# Diffraction limited centimeter scale radiator: metasurface grating antenna for phased array LiDAR

**WEIQIANG XIE**[1,2], **JINXI HUANG**[1,2*], **TIN KOMLJENOVIC**[1], **LARRY COLDREN**[1], **JOHN BOWERS**[1]

[1]*Electrical and Computer Engineering Department, University of California, Santa Barbara, Santa Barbara, California, 93106, USA*
[2]*J.H and W.X contributed equally*
*\*Corresponding author: jinxihuang@ucsb.edu*

**Abstract:** In this paper, we demonstrate a radiation-strength-chirped metasurface centimeter-long grating antenna, whose beam quality is approaching the diffraction limit in its footprint with 90% efficiency. Instead of chirping the duty cycle to tune radiation strength, which calls for non-scalable fabrication resolution, we design a 2D fish-bone surface grating with apordized width of the "bone". The meta surface is made by silicon nitride on top of silicon waveguide to weakly perturb the mode. Moreover, a Monte Carlo Markov Chain(MCMC) algorithm is implemented to design the radiation strength profile, making the optimized trade off between efficiency, beam quality and fabrication challenge. A metal mirror is designed to deposit on the top of the cladding for unidirectional upside-down radiation and uniform lateral gain. For the 1cm grating, our design is for a 0.01°beam with 37000 1D directional gain and 90% efficiency, with 200nm feature size. Finally, we fabricated the antenna and measured the beam quality with Fourier imaging system, which gives result of 0.02° -10dB beam divergence. To the best of our knowledge, this is the state-of-the-art on-chip antenna with CMOS- foundry-compatible technology.

## 1. Introduction

Silicon photonics[2] success enables the potential integrated optical phased array (OPA) a bright future. Light Detection and Range(LiDAR), comes to be an intriguing research issue[1] owing to their applications in autonomous driving as well as aircraft 3D mapping.. C band is favored in future LiDAR map thanks to its eye safe power level and extraordinary penetration in atmosphere. For OPA LiDAR, grating antenna is the most prominent candidate[6] of antenna thanks to its perfect compatibility with planar CMOS fabrication and high fabrication error tolerance.

In an OPA, as shown in Fig.1(a), the grating antennas lay parallel, with individual phase and thus beam steering on the dimension φ. Meanwhile, wavelength is tuned to steer the beam in θ dimension(Fig.1(b)), which is indicated by grating emission function

$$\sin\theta = n_{eff} - \lambda/\Lambda \qquad (1)$$

Where $n_{eff}$ is the effective refractive index of the waveguide fundamental mode and $\Lambda$ is the period of the grating. The larger scale the phased array is (ie, more elements and longer antenna effective length), the higher directivity we can acquire.

In this paper, a two dimensional metasurface pattern is proposed for the grating antenna to pursue diffraction limited effective aperture. Fig.1(d) shows the top view fish bone grating, whose grating groove is partially etched on the two sides. The radiation strength is chirped by apordizing width of the "bone". This paper is organized as follows. First, we show the simulation results of grating strength versus bone width and wavelength, which will be applied to design the metasurface. Comparably, the modulated grating strength by duty-cycle chirp is shown and analyzed. Next, we demonstrate the MCMC optimization to choose the best radiation strength profile and correspondingly the bone width profile, for the trade off between efficiency, beam quality and fabrication challenge. In addition, the position of gold mirror to grating L, which is deposited on the top of the $SiO_2$ cladding to achieve both unidirectional radiation as well as uniform gain in φ, is optimized to minimize directivity in φ. Finally, experiment is conducted to enhance our design.

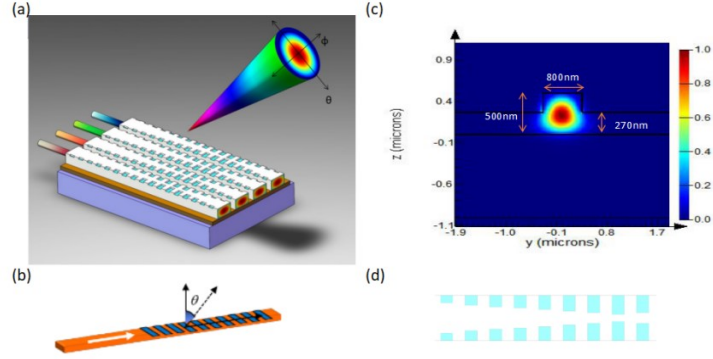

Fig. 1. (a) Optical phased array and grating antenna. Grating antennas lay parallel, of which the phase are controlled individually and thus interference for beam steering on the dimension φ. (b) Meanwhile, wavelength is tuned to steer the beam in θ. Waveguide parameter and fundamental mode is shown in (c). (d) shows the top view of the grating surface, blue windows are the etched groove

## 2. Methods and design

Our silicon photonics platform contains 1um buried silicon oxide(BOX) and 500nm top silicon, with rib waveguide whose stripe thickness is 230nm. In our grating antenna, the rib waveguide is determined due to low loss and single TE mode(Fig. 1(c)); meanwhile waveguide width is optimized to 800nm in order to minimize mode diameter to ensure both lower crosstalk between antennas in array and roughness scattering. Grating period is 560nm for radiation direction 23.6° at 1550nm.

The radiation angle tuning efficiency is given by the derivative of the grating function, that is,

$$\begin{aligned}\frac{d\theta}{d\lambda} &= (\frac{dn_{eff}}{d\lambda} - 1/\Lambda)/\cos\theta \\ &= (\frac{n_{eff} - n_g}{\lambda} - \frac{n_{eff}}{\lambda})/\cos\theta + \tan\theta/\lambda \\ &= (\sin\theta - n_g)/\lambda\cos\theta\end{aligned} \quad (3)$$

Putting the group index of our waveguide into the function, we find that it is 0.14°/nm. Additionally, 560nm period is chosen as the trade off between feature size and radiation efficiency.

There have been plenty of researches[3][4][7] conducted to optimize the element scale and position, along with phase for higher directivity and wider field of view in φ. However, beam optimization in θ is rarely concerned. Slower radiation rate for longer effective length, and thus sharper beam is necessary, while simply weakening the grating strength[5] is far from enough since the physical footprint is not exploited due to the exponential decay of radiated power along the grating. Previous work[6] tried duty cycle chirp for uniformly radiated power along the grating for larger effective length approaching the physical length. Nevertheless, to satisfy high antenna efficiency (e.g, >90%), the radiation strength R needs to cover a wide range ($R_{max}/R_{min}$=10 is necessary). Thus feature size at tens of nm is required in that approach, calling for nonscalable fab technology, like E-beam lithography. What's more, a uniform emission pattern leads to sidelobes due to truncation effect at the two ends of the antenna. Finally, the radiation strength of a duty cycle chirped grating is sensitive to wavelength and consequently ripple of radiation power at different wavelength as shown in Fig. 2. Another approach is sidewall grating[7], by modulating the width of sidewall teeth. However, controlling the teeth thickness with nm precision is challenging as well. As a consequence, fish-bone metsurface grating stands out thanks to its wide range of radiation strength modulation without the need of small feature size as well as its tolerance to lithography error.

Along with radiation strength chirping, for the sake of bone-width-apordization arised $n_{eff}$ varying, we taper the width of the waveguide simultaneously to leave the effective index unchanged at any position, and consequently maintain thebeam divergence at different scattering direction θ. The maximum change of waveguide width is 16nm, which corresponds to the bone width changing from 200nm to 400nm. Noticeably, chirping period to maintain θ[7] is not acceptable, since different grating period leads to different θ tuning efficiency, and thus deterioration of beam when wavelength is shifted. For duty-cycle chirp, the same problem exists.

The grating strength vs "bone" width and wavelength is shown in Fig.2(a). Silicon nitride depositing on the top of waveguide plays the role of the grating teeth due to their weak perturbation and selectional etching.The nitride teeth thickness is 60nm. Paralleled with fishbone grating, duty cycle chirp grating is shown in Fig. 2(b). As can be seen, duty cycle chirp grating calls for 50nm feature size to get $R_{max}/R_{min}=6$, which can exploit the footprint of grating for large effective aperture and will be mentioned in next paragraph. However, 120nm feature size is satisfactory for fishbone grating to get $R_{max}/R_{min}=6$, let alone its advantage of wavelength-insensitive radiation strength. This phenomenon is attributed to the principle behind grating radiation. Periodically perturbed index leads to periodical electric displacement $\Delta D(x)=\Delta\varepsilon(x)E$, which can be treated as volume current(VC) radiation source as shown in Payne's work[1] with $J_e=j\omega\Delta D$. From antenna theory, far field radiation will be the Fourier transformation of the VC multiplying dipole's directional gain, whose radiative component is that with $|k_x|<k_0$. In periodical structure, for example grating, radiation strength will be proportional to the first order Fourier series of the VC. Fishbone metasurface takes advantage of field non-uniformity in waveguide mode, as Fig. 1(c) shows, thus grating strength modulation efficiency is greatly increased when etching window gets to be larger. Consequently, tiny feature size(<100nm) is not necessary for low grating strength. Nevertheless, the duty cycle chirp linearly tunes first order Fourier component which results in the necessity of 60nm feature size for low grating strength. Moreover, the mode spreads out in relation to wavelength increment. The more the VC is close to the center of the mode, the more the VC intensity will be changed by wavelength tuning. This gives rise to the better bandwidth of the fishbone grating .

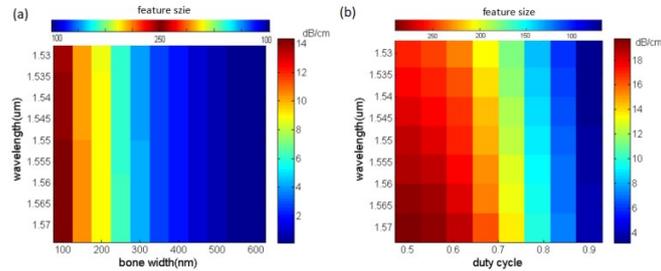

Fig. 2. Grating strength vs geometric parameters of grating. (a) shows the map of 2D fishbone grating, (b) shows the duty cycle chirp grating

As mentioned before, simply extending the grating effective length to get narrow beam by uniform radiation, we will suffer from fabrication challenge due to the requirement of small feature size, and sacrifice sidelobe suppression along with main beam power. Based on the property of truncation, the radiated power at the two ends of the grating should be weak to reduce sidelobes. Meanwhile, the effective aperture should be maximized with this constraint. Heuristically, the radiation field intensity f(x) along the waveguide following Gaussian profile will be a good choice, as shown in Fig.3(a). So we set the radiation profile to be two half-Gaussians connected together, with the peak located at position u, HWHM of the first gaussian $\sigma_1$ and the second $\sigma_2$. Power remaining in waveguide at any point X is expressed as

$$P(X) = 1 - \int_X f(x)dx = \begin{cases} 1 - (\int_0^u AG(\frac{x-u}{\sigma_1})dx + \int_u^X AG(\frac{x-u}{\sigma_2})dx) & X > u \\ 1 - \int_0^X AG(\frac{x-u}{\sigma_1})dx & X < u \end{cases}$$

where A is amplitude, $G(x) = \exp(-x^2/2)$.

When X=1cm, P=0.1 as we set antenna efficiency 90%. Correspondingly, grating strength profile is

$$R(x) = \frac{f(x)}{1 - \int_x f(x)dx}$$

The far field F ($\theta$) is

$$F(\theta) = (1/\cos^2(\theta)) \times F^2[\sqrt{f(x/\lambda)}]_{(\sin\theta)}$$

where F[f(x)]$_{(\sin\theta)}$ is a Fourier transform.

Therefore, beam directivity is non-analytically written as

$$directivity = \frac{\pi \max(F)}{\int F(\theta)d\theta}$$

Fabrication challenge is the ratio of C=$R_{max}/R_{min}$. The larger the C is, the more penalty we have. Thus the fab penalty is defined as

$$Penalty = \begin{cases} (C-2.5)^2 & 10 > R\max/R\min > 2.5 \\ 0 & R\max/R\min < 2.5 \\ \text{infinity} & R\max/R\min > 10 \end{cases}$$

The aimed function to be optimized is as follows

$$Score(\sigma_1, \sigma_2, u, A) = \alpha directivity + \beta sidelobesuppression - \gamma Pen \qquad (2)$$

constraint to P(1cm)=0.1

Apparently, this is an extremely non-explicit and non-convex optimization, with non-convex boundary. So the gradient-descent algorithm is not available here. Once we can learn the map of solution space of score function, optimization can be achieved as well. In machine learning, the Markov Chain Monte Carlo method is developed for Bayesian inference learning[2], which is available here. If we normalize score as a probability space, sampling following a Markov Chain of the space will have highest density in maximal point of score. Once we anneal the sampling process, we converge there. Here we propose a sampling transfer function from step n to n+1, which is Markovian in the space:

$$P(n \to n+1) = \begin{cases} 1 & if Score(n+1) > Score(n) \\ k(n)\frac{score(n+1)}{score(n)} & else \end{cases}$$

This is called Gibbs sampling. We anneal k, and parameters will converge to the global minimum of score.

The converging curve of optimization is shown in Fig. 3(d). After optimization, we acquire the parameter setup, and the grating strength curve is shown in Fig. 3(c), whose feature size is 200nm with regard to $C=R_{max}/R_{min}=4$. The 1D directional gain is 37,000. The -10dB beam width is 0.02°. Comparatively, we show the strength curve of uniform strength grating and the optimized strength profile of duty cycle chirped grating with 120nm feature size (whose C=2.4). It can be seen that beam divergence is relatively constant once we optimize the strength profile given a certain fab limit and efficiency. That is, the effective aperture is relatively constant after optimization. However, a tradeoff happens between fab challenge and directivity. As the comparison between the green and red plot, we notice that lower $R_{max}/R_{min}$ leads to relatively strong sidelobe near the beam, thus the peak intensity of far field of the C=2.4 grating is 5% weaker than that of C=4. In the score function, if we give more weight to fab challenge, optimized C will be smaller and vice versa. Meanwhile, it's apparent from the blue curve that naively pursuing a large aperture by uniform intensity sacrifices fab challenge while get almost nothing.

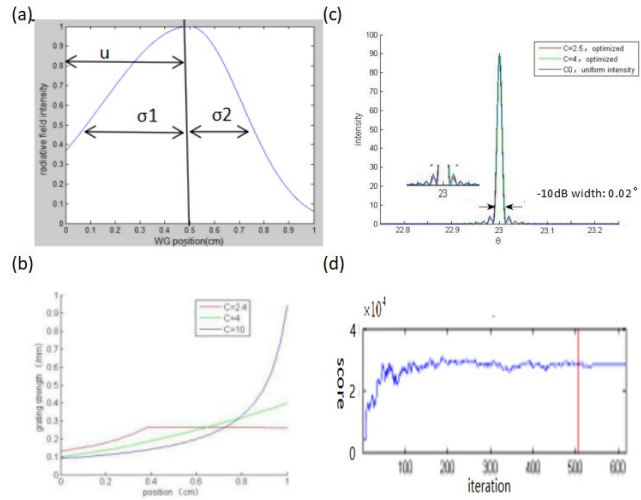

Fig. 3. Based on 90% efficiency. (a) shows our radiative near field setup, with two partial gaussian connected. (b) shows the optimized grating strength along grating: the curve C=4 is our optimized result; curve C=2.4 shows the optimized result when $R_{max}/R_{min}$=2.4, corresponding to duty cycle chirped grating with tiniest feature 120nm; curve C=10 shows the uniform near field grating. (c) shows the far field beam of the three strength profile. (d) is the converging curve of optimization.

Finally, to employ a grating antenna in a phased array, some other challenges must be solved. Above all, bidirectional radiation (symmetrically up and down) reduced the radiated power and detection efficiency since LiDAR applications require unidirectional beam. Thus, gold metal is deposited on the top of cladding as the mirror for TE mode reflection[10] and consequently upside-down unidirectional radiation. Additionally, in the φ direction, gain of single grating antenna is not uniform; instead, it is tapered by directional gain of dipole (cosφ). For the sake of wider steering range in φ, our antenna design should pursue lower gain in φ. In our application, FOV is ±40°. A simplified model for φ gain with metal mirror is a dipole placed in front of the metal plane. The normalized gain function is

$$f(\varphi) = F(\varphi)/\max(F(\varphi)),$$
$$F(\varphi) = |(1+\exp(2\pi j L \sqrt{\cos^2(\varphi)-\sin^2(\theta)}/\lambda))\cos(\varphi)|^2 \quad (3)$$

where φ is the direction of beam radiation, L is the distance between mirror and dipole. L should be larger than 1um thus the metal won't perturb the evanescent field of the grating mode and avoid introducing unexpected loss. A more precise gain in φ and radiation strength is acquired directly from FDTD simulation with far field monitor beneath the substrate, since cavity effect affects radiation strength[10]. When we sweep the distance as shown in Fig. 4(c), it's noticeable that FWHM increases; however, when L>1.12um, the central valley becomes lower than 0.75. Thus L=1.11um is chosen. This mirror simultaneously achieves unidirectional radiation and wide beam sweeping range. Our simulation shows that only 5% to 6% of the radiated power is absorbed by the gold metal depending on wavelength.

Since the beam is emitted from the bottom of the substrate, antireflection(AR) coating for wide incident angle range θ is necessary. Transmission line model is employed to design the AR coating, optimizing the material and thickness of layers. Available materials include alumina, silica and silicon nitride. To simplify fabrication, 5 layers are applied.

$$R = \sum_{\substack{\theta=0 \\ \Delta\theta=10}}^{50} \frac{Z(\theta,\vec{d},\vec{n}) - Z_s(\theta)}{Z(\theta,\vec{d},\vec{n}) + Z_s(\theta)} \tag{4}$$

R is the reflection summation with incident angle 0,10,20,30,40,50°, where Z is impedance of transmitting wave, d is the thickness vector of layers and n are their refractive indices. After optimization of d and n, the layer parameters are listed in table in the end, with transmission vs angle plotted in Fig. 4(b). Note that transmission at all direction is higher than 99%.

**Table. AR coating**

| layer | SiO$_2$ | TiO$_2$ | SiO$_2$ | TiO$_2$ | SiO$_2$ |
|---|---|---|---|---|---|
| Thickness(nm) | 30 | 118 | 354 | 274 | 213 |

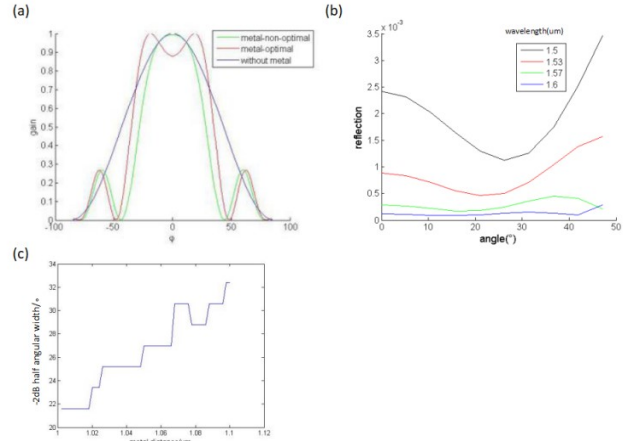

Fig. 4. (a) shows the lateral directional gain with mirror at optimized position, without mirror, and mirror placed at 1um to grating. (b) shows the reflection of substrate AR coating. (c) shows the relationship between mirror position and half angular width with central valley>0.8.

## 3. Experiment

We fabricate the grating antenna with a DUV stepper 3. 60nm nitride grating teeth deposited with PECVD and is dry etched to form the teeth. The grating metasurface is shown in Fig. 5(b). To simplify measurement of beam quality of the antenna, no top mirror is applied. The beam is measured with a Fourier imaging system developed in [9]. The system has 0.0067° resolution and omnidirectional FOV. The beams with different wavelength are shown in Fig.5(a), occupying 3 pixels and thus 0.02°, which is -10dB beam divergence. Weak noise be attributed to the higher loss rate due to sidewall roughness. A potential problem is the 30% sidelobe close to the mainlobe as shown in Fig. 5(c). We will deal with the principle behind sidelobe and potential solution in future work. Tuning efficiency is 0.14°/nm. Radiation loss is 10dB, occupying 90% power of incidence Fig. 5(d), matching our design pursuit.

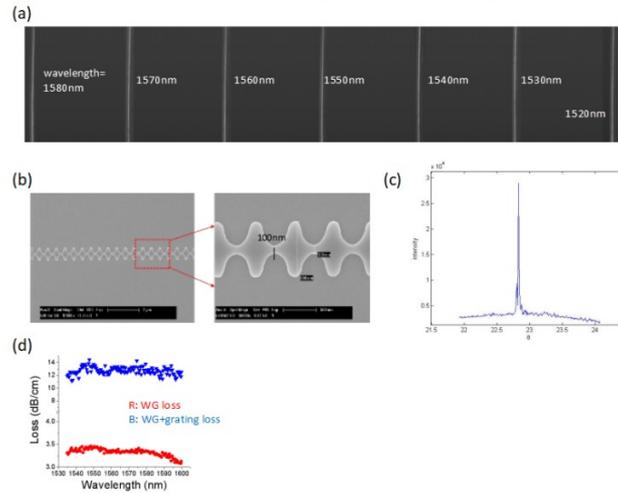

Fig. 5. Experimental measurement. (a) shows the beam (b) shows the SEM of metasurface grating (c) shows the beam profile at1540nm (d) shows the transmission loss of waveguide and grating

## 4. Conclusion

In this paper, we explore the optimization for grating antenna to get state-of-the-art on-chip beam quality with high sidelobe suppression and diffraction limited directivity. A fishbone grating with nitride teeth is used for wide radiation rate tuning range. MCMC optimization is executed to optimized radiation profile. The performance of the designed grating is robust with fab error of foundry scalable technology. Unidirectional radiation is explored by top mirror along with uniform gain in lateral radiation. This element will be an essential issue in integrated OPA LiDARs.

## Funding

DARPA MTO (MOABB HR0011-16-C-0106)

## Acknowledgement

The authors thank Paul Suni and James R. Colosimo from Lockheed Martin for useful discussions. This research was developed with funding from the Defense Advanced Research Projects Agency (DARPA). The views, opinions and/or findings expressed are those of the author and should not be interpreted as representing the official views or policies of the Department of Defense or the U.S. Government.